\documentstyle[12pt,graphicx,amsfonts]{article}

\def\fracov#1#2{{#1\over #2}}

\newcommand{\beq}{\begin{equation}}
\newcommand{\eeq}{\end{equation}}
\newcommand{\bea}{\begin{eqnarray}}
\newcommand{\eea}{\end{eqnarray}}
\newcommand{\ena}{\end{eqnarray}}

\newcommand{\Sb}{\Bar{\Sigma}}
\newcommand{\Db}{\Bar{D}}
\def\mathfrak{\bf}
\newcommand {\non}{\nonumber}

\renewcommand{\(}{\left(}
\renewcommand{\)}{\right)}
\renewcommand{\[}{\left[}
\renewcommand{\]}{\right]}

\newcommand{\sba}{\Bar{\sigma}}

\def\be{\begin{equation}}
\def\ee{\end{equation}}
\def\bea{\begin{eqnarray}}
\def\eea{\end{eqnarray}}
\def\dt#1{\on{\hbox{\bf .}}{#1}}                % (big) dot over
\def\Dot#1{\dt{#1}}
\def\IR{\relax{\rm I\kern-.18em R}}
\def\binomial#1#2{\left(\,{\buildrel
{\raise4pt\hbox{$\displaystyle{#1}$}}\over
{\raise-6pt\hbox{$\displaystyle{#2}$}}}\,\right)}

\def\[{\lfloor{\hskip 0.35pt}\!\!\!\lceil}
\def\]{\rfloor{\hskip 0.35pt}\!\!\!\rceil}

\newcommand{\AmS}{{\protect\the\textfont2
  A\kern-.1667em\lower.5ex\hbox{M}\kern-.125emS}}

\catcode`@=11
\def\un#1{\relax\ifmmode\@@underline#1\else
        $\@@underline{\hbox{#1}}$\relax\fi}
\catcode`@=12

\def\ad{{\kern0.5pt
                   \alpha \kern-5.05pt
\raise5.8pt\hbox{$\textstyle.$}\kern
0.5pt}}

\def\Dot#1{{\kern0.5pt
     {#1} \kern-5.05pt \raise5.8pt\hbox{$\textstyle.$}\kern
0.5pt}}

\def\a{\alpha}
\def\b{\beta}
\def\c{\chi}
\def\d{\delta}

\def\g{\gamma}

\def\l{\lambda}

\def\s{\sigma}

\def\z{\zeta}

\def\P{\Pi}

\def\S{\Sigma}

\def\cn{{\cal N}}

\def\cu{{\cal U}}
\def\cv{{\cal V}}

\def\bo{{\raise.15ex\hbox{\large$\Box$}}}               % D'Alembertian
\def\pa{\partial}                                       % curly d
\def\TH{{\raise.2ex\hbox{$\displaystyle \bigodot$}\mskip-4.7mu \llap H
\;}}
\def\face{{\raise.2ex\hbox{$\displaystyle \bigodot$}\mskip-2.2mu \llap
{$\ddot
        \smile$}}}                                      % happy face
\def\Bar#1{\overline{#1}}                       % big bar
\def\leftrightarrowfill{$\mathsurround=0pt \mathord\leftarrow \mkern-6mu
        \cleaders\hbox{$\mkern-2mu \mathord- \mkern-2mu$}\hfill
        \mkern-6mu \mathord\rightarrow$}
\def\dvec#1{\vbox{\ialign{##\crcr
        \leftrightarrowfill\crcr\noalign{\kern-1pt\nointerlineskip}
        $\hfil\displaystyle{#1}\hfil$\crcr}}}           % <--> accent
\def\dt#1{{\buildrel {\hbox{\LARGE .}} \over {#1}}}     % dot-over forsp/sb
\def\frac#1#2{{\textstyle{#1\over\vphantom2\smash{\raise.20ex
        \hbox{$\scriptstyle{#2}$}}}}}                   % fraction
\def\sfrac#1#2{{\vphantom1\smash{\lower.5ex\hbox{\small$#1$}}\over
        \vphantom1\smash{\raise.4ex\hbox{\small$#2$}}}} % alternatefraction
\def\bfrac#1#2{{\vphantom1\smash{\lower.5ex\hbox{$#1$}}\over
        \vphantom1\smash{\raise.3ex\hbox{$#2$}}}}       % "
\def\afrac#1#2{{\vphantom1\smash{\lower.5ex\hbox{$#1$}}\over#2}}    % "
\def\on#1#2{\mathop{\null#2}\limits^{#1}}               % arbitraryaccent
\newskip\humongous \humongous=0pt plus 1000pt minus 1000pt

\newif\ifdtup

  \def\pp{{\mathchoice
          {%displaystyle
              \kern 1pt%
              \raise 1pt
              \vbox{\hrule width5pt height0.4pt depth0pt
                    \kern -2pt
                    \hbox{\kern 2.3pt
                          \vrule width0.4pt height6pt depth0pt
                          }
                    \kern -2pt
                    \hrule width5pt height0.4pt depth0pt}%
                    \kern 1pt
           }
            {%textstyle
              \kern 1pt%
              \raise 1pt
              \vbox{\hrule width4.3pt height0.4pt depth0pt
                    \kern -1.8pt
                    \hbox{\kern 1.95pt
                          \vrule width0.4pt height5.4pt depth0pt
                          }
                    \kern -1.8pt
                    \hrule width4.3pt height0.4pt depth0pt}%
                    \kern 1pt
            }
            {%scriptstyle
              \kern 0.5pt%
              \raise 1pt
              \vbox{\hrule width4.0pt height0.3pt depth0pt
                    \kern -1.9pt  %[e]=0.15pt
                    \hbox{\kern 1.85pt
                          \vrule width0.3pt height5.7pt depth0pt
                          }
                    \kern -1.9pt
                    \hrule width4.0pt height0.3pt depth0pt}%
                    \kern 0.5pt
            }
            {%scriptscriptstyle
              \kern 0.5pt%
              \raise 1pt
              \vbox{\hrule width3.6pt height0.3pt depth0pt
                    \kern -1.5pt
                    \hbox{\kern 1.65pt
                          \vrule width0.3pt height4.5pt depth0pt
                          }
                    \kern -1.5pt
                    \hrule width3.6pt height0.3pt depth0pt}%
                    \kern 0.5pt%}
            }
        }}

  \def\mm{{\mathchoice
                       {%displaystyle
                             \kern 1pt
               \raise 1pt    \vbox{\hrule width5pt height0.4pt depth0pt
                                  \kern 2pt
                                  \hrule width5pt height0.4pt depth0pt}
                             \kern 1pt}
                       {%textstyle
                            \kern 1pt
               \raise 1pt \vbox{\hrule width4.3pt height0.4pt depth0pt
                                  \kern 1.8pt
                                  \hrule width4.3pt height0.4pt depth0pt}
                             \kern 1pt}
                       {%scriptstyle
                            \kern 0.5pt
               \raise 1pt
                            \vbox{\hrule width4.0pt height0.3pt depth0pt
                                  \kern 1.9pt
                                  \hrule width4.0pt height0.3pt depth0pt}
                            \kern 1pt}
                       {%scriptscriptstyle
                           \kern 0.5pt
             \raise 1pt  \vbox{\hrule width3.6pt height0.3pt depth0pt
                                  \kern 1.5pt
                                  \hrule width3.6pt height0.3pt depth0pt}
                           \kern 0.5pt}
                       }}

\def\pd{{\kern0.5pt
                   + \kern-5.05pt \raise5.8pt\hbox{$\textstyle.$}\kern
0.5pt}}

\def\pmd{{\kern0.5pt
                  \pm \kern-5.05pt \raise6.3pt\hbox{$\textstyle.$}\kern1.5pt}}

\def\md{{\mathchoice
   {%displaystyle
      {{\kern 1pt - \kern-6.2pt \raise5pt\hbox{$\textstyle.$}\kern 1pt}}}
    {%textstyle
      {{\kern 1pt - \kern-6.2pt \raise5pt\hbox{$\textstyle.$}\kern 1pt}}}
    {%scriptstyle
      {\kern0.5pt - \kern-5.05pt \raise3.4pt\hbox{$\textstyle.$}\kern0.5pt}}
    {%scriptscriptstyle
      {\kern0.5pt - \kern-5.05pt \raise3.4pt\hbox{$\textstyle.$}\kern0.5pt}}}}

\def\ad{{\dot{\alpha}}}
\def\bd{{\dot{\beta}}}

\def\pp{{\mathchoice
          {%displaystyle
              \kern 1pt%
              \raise 1pt
              \vbox{\hrule width5pt height0.4pt depth0pt
                    \kern -2pt
                    \hbox{\kern 2.3pt
                          \vrule width0.4pt height6pt depth0pt
                          }
                    \kern -2pt
                    \hrule width5pt height0.4pt depth0pt}%
                    \kern 1pt
           }
            {%textstyle
              \kern 1pt%
              \raise 1pt
              \vbox{\hrule width4.3pt height0.4pt depth0pt
                    \kern -1.8pt
                    \hbox{\kern 1.95pt
                          \vrule width0.4pt height5.4pt depth0pt
                          }
                    \kern -1.8pt
                    \hrule width4.3pt height0.4pt depth0pt}%
                    \kern 1pt
            }
            {%scriptstyle
              \kern 0.5pt%
              \raise 1pt
              \vbox{\hrule width4.0pt height0.3pt depth0pt
                    \kern -1.9pt  %[e]=0.15pt
                    \hbox{\kern 1.85pt
                          \vrule width0.3pt height5.7pt depth0pt
                          }
                    \kern -1.9pt
                    \hrule width4.0pt height0.3pt depth0pt}%
                    \kern 0.5pt
            }
            {%scriptscriptstyle
              \kern 0.5pt%
              \raise 1pt
              \vbox{\hrule width3.6pt height0.3pt depth0pt
                    \kern -1.5pt
                    \hbox{\kern 1.65pt
                          \vrule width0.3pt height4.5pt depth0pt
                          }
                    \kern -1.5pt
                    \hrule width3.6pt height0.3pt depth0pt}%
                    \kern 0.5pt%}
            }
        }}

  \def\mm{{\mathchoice
                       {%displaystyle
                             \kern 1pt
               \raise 1pt    \vbox{\hrule width5pt height0.4pt depth0pt
                                  \kern 2pt
                                  \hrule width5pt height0.4pt depth0pt}
                             \kern 1pt}
                       {%textstyle
                            \kern 1pt
               \raise 1pt \vbox{\hrule width4.3pt height0.4pt depth0pt
                                  \kern 1.8pt
                                  \hrule width4.3pt height0.4pt depth0pt}
                             \kern 1pt}
                       {%scriptstyle
                            \kern 0.5pt
               \raise 1pt
                            \vbox{\hrule width4.0pt height0.3pt depth0pt
                                  \kern 1.9pt
                                  \hrule width4.0pt height0.3pt depth0pt}
                            \kern 1pt}
                       {%scriptscriptstyle
                           \kern 0.5pt
             \raise 1pt  \vbox{\hrule width3.6pt height0.3pt depth0pt
                                  \kern 1.5pt
                                  \hrule width3.6pt height0.3pt depth0pt}
                           \kern 0.5pt}
                       }}

\def\pd{{\kern0.5pt
                   + \kern-5.05pt \raise5.8pt\hbox{$\textstyle.$}\kern
0.5pt}}

\def\pmd{{\kern0.5pt
                  \pm \kern-5.05pt \raise6.3pt\hbox{$\textstyle.$}\kern1.5pt}}

\def\md{{\mathchoice
   {%displaystyle
      {{\kern 1pt - \kern-6.2pt \raise5pt\hbox{$\textstyle.$}\kern 1pt}}}
    {%textstyle
      {{\kern 1pt - \kern-6.2pt \raise5pt\hbox{$\textstyle.$}\kern 1pt}}}
    {%scriptstyle
      {\kern0.5pt - \kern-5.05pt \raise3.4pt\hbox{$\textstyle.$}\kern0.5pt}}
    {%scriptscriptstyle
      {\kern0.5pt - \kern-5.05pt \raise3.4pt\hbox{$\textstyle.$}\kern0.5pt}}}}

\def\dslash{\not{\hbox{\kern-2pt $\partial$}}}
\def\Dslash{\not{\hbox{\kern-4pt $D$}}}
\def\pslash{\not{\hbox{\kern-2.3pt $p$}}}
 \newtoks\slashfraction
 \slashfraction={.13}
 \def\slash#1{\setbox0\hbox{$ #1 $}
 \setbox0\hbox to \the\slashfraction\wd0{\hss \box0}/\box0 }

\font\ro=cmsy10                          % font with rope
\def\kcr{{\hbox{\ro \char'170}}}                % right-handed rope
\def\ktl{{\hbox{\ro \char'170}}}        % top end for left-handed rope
\def\ktr{{\hbox{\ro \char'170}}}        % " right
\def\kbl{{\hbox{\ro \char'170}}}        % " bottom left
\def\kbr{{\hbox{\ro \char'170}}}        % " right
\def\plpl{\raise-2pt\hbox{$\raise3pt\hbox{$_+$}\hskip-6.67pt\raise0.0pt
\hbox{$^+$}\hskip 0.01pt$}}
\def\mimi{\raise-2pt\hbox{$\raise3pt\hbox{$_-$}\hskip-6.67pt\raise0.0pt
\hbox{$^-$}\hskip 0.01pt$}}

\def\bo{{\raise.15ex\hbox{\large$\Box$}}}               % D'Alembertian
\def\pa{\partial}                                       % curly d
\def\TH{{\raise.2ex\hbox{$\displaystyle \bigodot$}\mskip-4.7mu \llap H \;}}
\def\face{{\raise.2ex\hbox{$\displaystyle \bigodot$}\mskip-2.2mu \llap {$\ddot
        \smile$}}}                                      % happy face
\def\Bar#1{\overline{#1}}                       % big bar
\def\leftrightarrowfill{$\mathsurround=0pt \mathord\leftarrow \mkern-6mu
        \cleaders\hbox{$\mkern-2mu \mathord- \mkern-2mu$}\hfill
        \mkern-6mu \mathord\rightarrow$}
\def\dvec#1{\vbox{\ialign{##\crcr
        \leftrightarrowfill\crcr\noalign{\kern-1pt\nointerlineskip}
        $\hfil\displaystyle{#1}\hfil$\crcr}}}           % <--> accent
\def\dt#1{{\buildrel {\hbox{\LARGE .}} \over {#1}}}     % dot-over for sp/sb
\def\frac#1#2{{\textstyle{#1\over\vphantom2\smash{\raise.20ex
        \hbox{$\scriptstyle{#2}$}}}}}                   % fraction
\def\sfrac#1#2{{\vphantom1\smash{\lower.5ex\hbox{\small$#1$}}\over
        \vphantom1\smash{\raise.4ex\hbox{\small$#2$}}}} % alternate fraction
\def\bfrac#1#2{{\vphantom1\smash{\lower.5ex\hbox{$#1$}}\over
        \vphantom1\smash{\raise.3ex\hbox{$#2$}}}}       % "
\def\afrac#1#2{{\vphantom1\smash{\lower.5ex\hbox{$#1$}}\over#2}}    % "
\def\on#1#2{\mathop{\null#2}\limits^{#1}}               % arbitrary accent
\parskip=\medskipamount                 % space between paragraphs (LaTeX)
\thicklines                         % thick straight lines for pictures (LaTeX)

\def\oldheadpic{                                % old UM heading
        \setlength{\unitlength}{.4mm}
        \thinlines
        \par
        \begin{picture}(349,16)
        \put(325,16){\line(1,0){4}}
        \put(330,16){\line(1,0){4}}
        \put(340,16){\line(1,0){4}}
        \put(335,0){\line(1,0){4}}
        \put(340,0){\line(1,0){4}}
        \put(345,0){\line(1,0){4}}
        \put(329,0){\line(0,1){16}}
        \put(330,0){\line(0,1){16}}
        \put(339,0){\line(0,1){16}}
        \put(340,0){\line(0,1){16}}
        \put(344,0){\line(0,1){16}}
        \put(345,0){\line(0,1){16}}
        \put(329,16){\oval(8,32)[bl]}
        \put(330,16){\oval(8,32)[br]}
        \put(339,0){\oval(8,32)[tl]}
        \put(345,0){\oval(8,32)[tr]}
        \end{picture}
        \par
        \thicklines
        \vskip.2in}
\def\oldtitle#1#2#3#4{\oldheadpic\begin{center}\vglue.5in{\large\bf #1}\\[.6in]
        {#2}\\[.1in] {\it Department of Physics and Astronomy}\\
        {\it University of Maryland, College Park, MD 20742}\\[.6in]
        Physics Publication \#{#3}\\ {#4}\\[1.5in] {\bf ABSTRACT}\\[.1in]
        \end{center} \begin{quotation}}                 % old title stuff
\def\oldTitle#1#2#3#4#5#6#7{\oldheadpic\begin{center} \vglue .4in
        {\large\bf #1}\\[.4in]
        {#2}\\[.1in] {\it Department of Physics and Astronomy}\\
        {\it University of Maryland, College Park, MD 20742}\\[.1in]
        {#3}\\[.1in] {\it {#4}}\\ {\it {#5}}\\[.4in]
        Physics Publication \#{#6}\\ {#7}\\[.5in] {\bf ABSTRACT}\\[.1in]
        \end{center} \begin{quotation}}                 % " for 2 authors
\def\border{                                            % border
        \setlength{\unitlength}{1mm}
        \newcount\xco
        \newcount\yco
        \xco=-21
        \yco=12
        \begin{picture}(140,0)
        \put(\xco,\yco){$\ktl$}
        \advance\yco by-1
        {\loop
        \put(\xco,\yco){$\kcr$}
        \advance\yco by-2
        \ifnum\yco>-240
        \repeat
        \put(\xco,\yco){$\kbl$}}
        \xco=158
        \yco=12
        \put(\xco,\yco){$\ktr$}
        \advance\yco by-1
        {\loop
        \put(\xco,\yco){$\kcr$}
        \advance\yco by-2
        \ifnum\yco>-240
        \repeat
        \put(\xco,\yco){$\kbr$}}
        \put(-20,13){\tiny **University of Maryland * Center for String and
         Particle  Theory* Physics Department***University of Maryland *Center
        for String and Particle  Theory** }
        \put(-20,-241.5){\tiny **University of Maryland * Center for String and
         Particle  Theory* Physics Department***University of Maryland *Center
        for String and Particle  Theory** }
        \end{picture}
        \par\vskip-8mm}
\def\bordero{                                           % alternate border
        \setlength{\unitlength}{1mm}
        \newcount\xco
        \newcount\yco
        \xco=-31
        \yco=12
        \begin{picture}(140,0)
        \put(\xco,\yco){$\ktl$}
        \advance\yco by-1
        {\loop
        \put(\xco,\yco){$\kclr$}
        \advance\yco by-2
        \ifnum\yco>-240
        \repeat
        \put(\xco,\yco){$\kbl$}}
        \xco=151
        \yco=12
        \put(\xco,\yco){$\ktr$}
        \advance\yco by-1
        {\loop
        \put(\xco,\yco){$\kcr$}
        \advance\yco by-2
        \ifnum\yco>-240
        \repeat
        \put(\xco,\yco){$\kbr$}}
        \put(-20,12){\ooo bacdefghidfghghdhededbihdgdfdfhhdheidhdhebaaahjhhdahba

hgdedge
   hgfdiehhgdigicba}
        \put(-20,-241.5){\ooo ababaighefdbfghgeahgdfgafagihdidihiidhiagfedhadbfd

ecdcdfa
   gdcbhaddhbgfchbgfdacfediacbabab}
        \end{picture}
        \par\vskip-8mm}
\def\headpic{                                           % UM heading
        \indent
        \setlength{\unitlength}{.4mm}
        \thinlines
        \par
        \begin{picture}(29,16)
        \put(165,16){\line(1,0){4}}
        \put(170,16){\line(1,0){4}}
        \put(180,16){\line(1,0){4}}
        \put(175,0){\line(1,0){4}}
        \put(180,0){\line(1,0){4}}
        \put(185,0){\line(1,0){4}}
        \put(169,0){\line(0,1){16}}
        \put(170,0){\line(0,1){16}}
        \put(179,0){\line(0,1){16}}
        \put(180,0){\line(0,1){16}}
        \put(184,0){\line(0,1){16}}
        \put(185,0){\line(0,1){16}}
        \put(169,16){\oval(8,32)[bl]}
        \put(170,16){\oval(8,32)[br]}
        \put(179,0){\oval(8,32)[tl]}
        \put(185,0){\oval(8,32)[tr]}
        \end{picture}
        \par\vskip-6.5mm
        \thicklines}
\def\title#1#2#3#4{\border\headpic {\hbox to\hsize{#4 \hfill UMDEPP #3}}\par
        \begin{center} \vglue .5in {\large\bf #1}\\[.6in]
        {#2}\\[.1in] {\it Department of Physics and Astronomy}\\
        {\it University of Maryland, College Park, MD 20742}\\[1.5in]
        {\bf ABSTRACT}\\[.1in] \end{center} \begin{quotation}}  % title stuff
\def\Title#1#2#3#4#5#6#7{\border\headpic
        {\hbox to\hsize{#7 \hfill UMDEPP #6}}\par
        \begin{center} \vglue .4in {\large\bf #1}\\[.4in]
        {#2}\\[.1in] {\it Department of Physics and Astronomy}\\
        {\it University of Maryland, College Park, MD 20742}\\[.1in]
        {#3}\\[.1in] {\it {#4}}\\ {\it {#5}}\\[.5in] {\bf ABSTRACT}\\[.1in]
        \end{center} \begin{quotation}}                 % " for 2 authors
\def\endtitle{\end{quotation}\newpage}                  % end title page

\def\qd{{\kern0.5pt
                   q \kern-5.05pt \raise5.8pt\hbox{$\textstyle.$}\kern
0.5pt}}

\begin{document}

\def\dt#1{\on{\hbox{\bf .}}{#1}}                % (big) dot over
\def\Dot#1{\dt{#1}}

\def\gfrac#1#2{\frac {\scriptstyle{#1}}
        {\mbox{\raisebox{-.6ex}{$\scriptstyle{#2}$}}}}
\def\gg{{\hbox{\sc g}}}
%\border\headpic 
{\hbox to\hsize{October, 2006 \hfill
{UMDEPP 06--055}}}
\par
{$~$ \hfill
{Bicocca--FT--06--17}}
\par
~~~
{$~$ \hfill {hep-th/0610333}}
\par

\setlength{\oddsidemargin}{0.3in}
\setlength{\evensidemargin}{-0.3in}
\begin{center}
\vglue .10in
{\Large\bf New massive supergravity multiplets}
\\[.35in]

S.\, James Gates, Jr.${}^\dag$\footnote{{\it E-mail address}: gatess@wam.umd.edu},
Sergei M. Kuzenko${}^\ddag$\footnote{{\it E-mail address}:  kuzenko@cyllene.uwa.edu.au} and
Gabriele Tartaglino-Mazzucchelli${}^\star$\footnote{{\it E-mail address}: 
gabriele.tartaglino@mib.infn.it\\
$~~~~~~$Address after November 15, 2006: School of Physics M013, The University of Western Australia,
\\$~~~~\,$ 
35 Stirling Highway, Crawley W.A. 6009, Australia.}
\\[0.3in]
${}^\dag${\it Center for String and Particle Theory\\
Department of Physics, University of Maryland\\
College Park, MD 20742-4111 USA}\\[0.1in]
${^\ddag}${\it School of Physics M013, The University of Western Australia\\
35 Stirling Highway, Crawley W.A. 6009, Australia}\\[0.1in]
%{\it {and}}\\[0.1in]
${}^\star${\it Dipartimento di Fisica, Universit\`a degli studi
Milano-Bicocca\\and INFN, Sezione di Milano-Bicocca, piazza della Scienza 3,
I-20126 Milano, Italy}\\[0.6in]

%{\bf ABSTRACT}\\[.01in]
\end{center}
\begin{quotation}
%{???}

\begin{abstract}
\baselineskip=14pt
\noindent
We present new off-shell formulations for the massive superspin-3/2
multiplet. In the massless limit, they reduce respectively to
the old minimal  ($n=-1/3$) and non-minimal ($n\neq -1/3, 0$) 
linearized formulations for 4D $\cn=1$ supergravity. 
Duality transformations,   which  relate the models constructed, 
are derived.
\end{abstract}
%${~~~}$ \newline ${~~~}$ \newline PACS: 04.65.+e, 11.15.-q, 11.30.Pb, 12.60.Jv

\endtitle

\setcounter{equation}{0}
\section{Introduction}
${}$Four-dimensional  $\cn=1$ supergravity exists 
in several off-shell incarnations.
They differ in the structure of their auxiliary fields and, 
as a consequence, in their matter couplings to supersymmetric matter.
It is an ancient tradition\footnote{It goes back to 1977 when 
the prepotential formulation 
for $\cn=1$ superfield supergravity was
\\
${~~~~}$ first developed
\cite{Siegel,CompensatoriConformi}.}
to label the off-shell  $\cn=1$ supergravity formulations 
by a parameter $n$, 
with its different values corresponding to the following 
supergravity versions:
(i) non-minimal ($n\neq -1/3, 0$) 
\cite{Breitenlohner,CompensatoriConformi,GatesJV};
(ii) old minimal ($n=-1/3$) \cite{WZ-old,old};
(iii) new minimal ($n=0$) \cite{new}.
Comprehensive reviews of these formulations 
can be found in \cite{SUPERSPACE,BuchbinderKuzenko}.
At the linearized level, there also exists 
a third minimal realization 
for the massless $(3/2,2)$  supermultiplet
\cite{BGLP1}, which is 
reminiscent of the new minimal formulation.
The three minimal formulations and the non-minimal series
turn out to comprise all possible ways to realize the irreducible massless 
superspin-3/2 multiplet as a gauge theory  of a real axial vector 
$H_a$ (gravitational superfield) and special compensator(s) \cite{GKP}.
Somewhat unexpectedly, 
a proliferation of off-shell formulations emerges 
in the massive case. 

On the mass shell, there is a unique way to realize 
the massive superspin-3/2 multiplet
(or massive graviton multiplet)
in terms of a  real (axial) vector superfield $H_a$. 
The corresponding equations 
\cite{Sok,BuchbinderKuzenko,BGLP1}
are:
\bea
\label{32irrepsp}
(\Box-m^2)H_{\a\ad}=0~, \quad
D^\a H_{\a\ad}=0~, \quad
\Bar D^{\dot\a} H_{\a\ad}=0 \quad
\Longrightarrow  \quad \pa^{\a\ad}H_{\a\ad}=0~.
\eea
It turns out that no action functional exists
to generate these equations 
if $H_{\a \ad}$ is the only dynamical variable \cite{BGLP1}.
However, such an action can be constructed if one allows for 
auxiliary superfields $\varphi$ with the property that the full mass shell 
is equivalent to the equations (\ref{32irrepsp}) together with $\varphi =0$.
Several supersymmetric models with the required properties have been
proposed \cite{BGLP1,GSS,BGKP}.
In particular, for each of the three minimal formulations
for linearized supergravity, massive extensions 
have been derived \cite{GSS,BGKP}. By applying superfield duality 
transformations to these theories, one generates three more models
\cite{BGKP} two of which originally appeared in \cite{BGLP1}.

The present paper continues the research initiated in 
\cite{BGLP1,GSS,BGKP}.
We propose new off-shell formulations for the massive superspin-3/2
multiplet. In particular, we derive two new massive extensions
of old minimal supergravity, 
which possess quite interesting properties,
as well as a massive extension of 
non-minimal supergravity.

\setcounter{equation}{0}
\section{Minimal supergravity multiplets and their massive extensions} 
\label{Massive Graviton Multiplets}

In this section, we review the linearized actions for the 
three minimal supergravity formulations,  and 
recall  their massive extensions proposed in \cite{GSS,BGKP}.
These massive actions possess nontrivial duals \cite{BGLP1,BGKP}, 
which are collected in the Appendix.

\subsection{Minimal supergravity multiplets}

Throughout this paper, we use
a reduced set \cite{GKP} of the superprojectors \cite{SG}
for the gravitational superfield $H_{\a\ad}$:
\bea
\P^L_{0}H_{\a\ad}&=&-{{1}\over{32}}{\pa_{\a\ad}\over\Box^2}\{ D^2,
\Db^2 \}\pa^{\b\bd}H_{\b\bd}~~,\\
\P^L_{1\over 2}H_{\a\ad}&=&
{ 1\over {16}}{\pa_{\a\ad}\over\Box^2}
D^\g\Db^2D_\g \pa^{\b\bd}H_{\b\bd}~~, \\
\P^T_{1\over 2}H_{\a\ad}&=&{ 1\over {48}}{\pa_{~\dot{\a}}^{\b}\over\Box^2}
\Big[D_\b \Db^2D^\g \pa_{(\a}^{~~\dot{\b}}H_{\g)\dot{\b}}+D_\a\Db^2D^\g
\pa_{(\b}^{~~\dot{\b}}H_{\g)\dot{\b}}\Big]~~, \\
\P^T_{1}H_{\a\ad}&=&{1\over{32}}{\pa_{~\dot{\a}}^{\b}\over\Box^2}\{
D^2, \Db^2\} \pa_{(\a}^{~~\dot{\b}}H_{\b)\dot{\b}}~~,\\
\P^T_{3\over 2}H_{\a\ad}&=&
-{1\over{48}}{\pa_{~\dot{\a}}^{\b}\over\Box^2}D^\g\Db^2
D_{(\g}\pa_\a^{~\dot{\b}}H_{\b)\dot{\b}}~~.
\eea
Here the superscripts $L$ and $T$ denote longitudinal and transverse
projectors, while the subscripts $0, 1/2, 1, 3/2$ stand for superspin.
Given a local linearized action functional of $H_{\a \ad}$, it can be expressed 
in terms of  superprojectors using 
the following identities:
\bea
D^\g\Db^2D_\g H_{\a\ad} &=&
-8 \Box \(\P^L_{1\over 2}+\P^T_{1\over 2}+\P^T_{3\over 2}\)H_{\a\ad}~~,
\label{id1}\\
\pa_{\a\ad}\pa^{\b\bd}H_{\b\bd}&=&
-2\Box (\P^L_{0} +\P^L_{1\over 2})H_{\a\ad}~~,\label{id2}\\
\[D_\a,\Bar{D}_\ad\]\[ D^\b,\Bar{D}^\bd\] H_{\b\bd}
&=&
8 \Box \(\P_{0}^{L} -3\,\P^T_{1\over 2}\) H_{\a\ad}~~,\label{id3}\\
\Box H_{\a\ad} &=&  \Box\(\P^L_{0}+\P^L_{1\over 2} +\P^T_{1\over 2}+\P^T_{1}
+\P^T_{3\over 2}\)H_{\a\ad}~,
\eea

The linearized action 
for old minimal (type I) supergravity is
\be
S^{({\rm I})} [H, \s] = \int d^8z \,
\Big\{
H^{\a\ad}\Box\Big(\fracov{1}{2} \P^T_{3\over 2}- \fracov{1}{3} \P^L_{0}\Big)H_{\a\ad}
-i(\s -\Bar \s ) \pa^{\a\ad}  H_{\a\ad}
- 3  \Bar \s  \s
\Big\}~.
\label{oldM}
\ee
Here $\s$ is the chiral compensator, $\overline{ D}_\ad \s =0$.

The linearized action 
for new minimal (type II) supergravity is
 \bea
S^{({\rm II})} [H, \cu] &=&\int d^8z\,
\Big\{H^{\a\ad}\Box\Big(\fracov 12\P^T_{3\over 2}-\P^T_{1\over 2}\Big)H_{\a\ad}
+\fracov{1}{2}\cu [D_\a,\Bar D_{\dot\a}]H^{\a\ad}
+\fracov{3}{2}\cu^2\Big\}~. ~~~~~
\label{II}
\eea
Here $\cu$  is the real linear compensator, ${\Bar D}^2 \cu =0$.

Type III supergravity is known at the linearized 
level \cite{BGLP1} only. The corresponding action is
  \bea
S^{({\rm III})} [H, \cu] &=&\int d^8z\,
\Big\{H^{\a\ad}\Box\Big(\fracov{1}{2}\P^T_{3\over 2}+\fracov{1}{3}\P^L_{1\over 2}\Big)H_{\a\ad}
+ \cu \pa_{\a\ad}  H^{\a\ad}
+\fracov{3}{2}\cu^2\Big\}~.
\label{III}
\eea 
Similarly to (\ref{II}),
here
$\cu$ the real linear compensator, 
${\Bar D}^2 \cu =0$.

\subsection{Massive extensions}

As demonstrated in  \cite{GSS,BGKP},  consistent massive extensions 
of the supersymmetric theories (\ref{oldM}), (\ref{II}) and (\ref{III})
can be obtained simply by adding mass terms for the gravitational superfield
and for a gauge potential associated with  the compensator, 
with the latter being treated as a gauge-invariant field strength.

Consider first the off-shell massive supergravity
multiplet  derived in \cite{GSS}.
The chirality constraint on the compensator $\s$ 
in (\ref{oldM}) can always be solved in terms an unconstrained 
real superfield \cite{VariantSG}:
\bea
\s = -{1\over 4} \Bar D^2 P~, \qquad \sba=-{1\over 4} D^2 P~, 
 \qquad \quad 
\Bar P = P~.
\label{constrIAreal}
\eea
Then, the massive extension of (\ref{oldM}), 
which was  proposed in \cite{GSS}, is 
\bea
S^{({\rm I})}_{\rm mass} [H, P] =
S^{({\rm I})} [H, \s]
- {1\over 2} m^2\int d^8z \,
\Big\{H^{\a\ad} H_{\a\ad}
-\fracov{9}{2} P^2 \Big\}~.
\label{IA}
\eea

The supergravity formulations  (\ref{II}) and (\ref{III}) 
involve the real linear compensator $\cu$.
The constraint on $\cu$ can be solved as follows \cite{Siegel2}:
\bea
\cu=D^\a \chi_\a+\Bar D_{\dot\a}\Bar \chi^{\dot\a}~,
\qquad \Bar D_\ad \chi_\a = 0~,
\nonumber
\eea
with $\chi_\a$ an unconstrained chiral spinor.
Adopting $\chi_\a $ and ${\bar \chi}_\ad$ as independent 
dynamical variables to describe the compensator, 
the new minimal model (\ref{II})
 possesses the  massive extension  \cite{BGKP}
\bea
S^{({\rm II})}_{\rm mass} [H, \chi] &=&
S^{({\rm II})} [H, \cu]  - {1\over 2} m^2\int d^8z\,
H^{\a\ad}H_{\a\ad}
+3m^2 \Big\{  \int d^6z \,  \chi^2
+{\rm c. c.} \Big\}~.~~~~~
\label{IIA}
\eea
Similarly, the type III model (\ref{III})
possesses the  following massive extension  \cite{BGKP} 
\bea
S^{({\rm III})}_{\rm mass} [H, \chi] &=&
S^{({\rm III})} [H, \cu]  - {1\over 2} m^2\int d^8z\,
H^{\a\ad}H_{\a\ad}
-9m^2 \Big\{  \int d^6z \,  \chi^2
+{\rm c.c.} \Big\}~.~~~~~
\label{IIIA}
\eea

\setcounter{equation}{0}
\section{New massive supergravity multiplets}

In the previous section we have reviewed several known formulations 
for the massive superspin-$3/2$ multiplet. 
They constitute  massive extensions
of the minimal supergravity formulations 
with $12+12$ off-shell degrees of freedom.
Now,  we  are going to obtain
a massive extension of the non-minimal supergravity
formulation with $20+20$ off-shell degrees of freedom. 
In the notation of \cite{BuchbinderKuzenko},
the linearized action for non-minimal supergravity  
\cite{CompensatoriConformi} is as follows:
\bea
S^{\rm NM}[H,\S]&=&
\int d^8z\Bigg[
-{1\over 16}\,H^{\alpha\dot{\alpha}}D^\b\Db^2D_\b H_{\alpha\dot{\alpha}}
+{n+1\over 8n}\,(\pa_{\a\ad}H^{\alpha\dot{\alpha}})^2\non\\
&&\,\,\,
+{n+1\over 32}\,([D_\a,\Db_\ad]H^{\alpha\dot{\alpha}})^2
-{(n+1)(3n+1)\over 4n}\,iH^{\alpha\dot{\alpha}}\pa_{\alpha\dot{\alpha}}(\S-\Sb)
\non\\
&&\,\,\,
-{3n+1\over 4}\,H^{\alpha\dot{\alpha}}(D_\a\Db_\ad\S-\Db_\ad D_\a\Sb)\non\\
&&\,\,\,
+{(3n+1)^2\over 4n}\,\Sb\S
+{9n^2-1\over 8n}\,(\S^2+\Sb^2)
\Bigg]=\non\\\non\\
&=&
\int d^8z\Bigg[
\,{1\over 2}H^{\alpha\dot{\alpha}}\Box\Big(
\Pi^T_{{3\over 2}}
+
{(n+1)^2\over 2n}\,\Pi^L_0
+{3n+1\over 2n}\,\Pi^L_{{1\over 2}}
-{3n+1\over 2}\,\Pi^T_{{1\over 2}}
\Big)H_{\alpha\dot{\alpha}}
\non\\
&&\,\,\,
-{(n+1)(3n+1)\over 4n}iH^{\alpha\dot{\alpha}}\pa_{\alpha\dot{\alpha}}
(\S-\Bar{\S})
+{(3n+1)^2\over 4n}\,\Bar{\S}\S
\non\\
&&\,\,\,
-{3n+1\over 4}\,H^{\alpha\dot{\alpha}}(D_\a\Db_\ad\S-\Db_\ad D_\a\Sb)
+{9n^2-1\over 8n}(\S^2+\Bar{\S}^2)
\Bigg]
~,~~~~~~~~~~~~
\label{NM}
\eea
Here $n\ne -1/3, 0$, and the compensator
$\S$ 
is  a complex linear superfield obeying the only 
constraint $\Db^2\S=0$. For simplicity, the parameter $n$ is chosen 
in (\ref{NM}) 
to be real, see  \cite{CompensatoriConformi,GatesJV} for 
the general  case of complex $n$.

${}$From the point of view of massive supergravity, 
the non-minimal formulation appears to be quite special.
It turns out that there is no consistent massive extension 
of the theory (\ref{NM}) obtained by adding mass terms
for the gravitational superfield
and for the gauge spinor  potential $\Psi_\a$ 
associated with  the non-minimal compensator
$\Bar \S=D^\a \Psi_\a$.\footnote{Note that, of course, 
the unconstrained superfield prepotential superfield $\Psi^{\a}$ 
is not chiral, unlike 
${}\,\,\,~~\,$ $\chi_{\a}$ in the new minimal case.}
This fact is in obvious contrast with 
the minimal supergravity formulations discussed in the previous 
section. 
We will come back to a discussion of these  
points in section 4.

\subsection{Massive extensions of old minimal supergravity}

To derive a massive extension of (\ref{NM}), one can try to employ 
the idea that the old minimal and non-minimal supergravity 
formulations are dually equivalent, see e.g. 
\cite{SUPERSPACE,BuchbinderKuzenko}
for  reviews. In order to apply duality considerations in the massive 
case, however, 
it is necessary to have an appropriate massive extension
of the action (\ref{oldM}). It turns out that  the formulation (\ref{IA})
is not well suited.

Therefore, as a first step,
let us actually derive a new massive extension of 
the old minimal supergravity formulation (\ref{oldM}).
As compared with  (\ref{IA}),  such an extension appears to 
be more natural, for the chiral compensator is 
defined through an  
{\it unconstrained complex} superfields $F$: 
\bea
\s=-{1\over 4}\Db^2 F ~,  \qquad \sba=-{1\over 4}D^2\Bar{F}~.
\label{constrIAcomplex}
\eea
We choose the simplest ansatz  for the massive action:  
\bea
\tilde{S}^{({\rm I})}_{\rm mass} [H, F] 
&=&
S^{({\rm I})} [H, \s]
- m^2
\int d^8z \,
\Big\{ {1\over 2} H^{\a\ad} H_{\a\ad}
-a 
F\Bar{F}
 \Big\}~,
\label{IAcomplex}
\eea
with  $a\neq 0$  a real constant. 

To prove that (\ref{IAcomplex}) indeed describes 
a massive superspin-$3/2$  multiplet, for a special value of
the parameter  $a$, we study the 
corresponding equations of motion: 
\bea
0&=&
\Box\Big(-{2\over 3}\Pi^L_0+\Pi^T_{{3\over 2}}\Big) H_{\alpha\dot{\alpha}}
+i\pa_{\alpha\dot{\alpha}}(\sigma-\Bar{\sigma})
-m^2H_{\alpha\dot{\alpha}}~,
\label{eqH1} \\&&\non\\
0&=&
{i\over 4}\Db^2\pa^{\alpha\dot{\alpha}}H_{\alpha\dot{\alpha}}
-{3\over 16}\Db^2D^2\Bar{F}+a\,m^2\Bar{F} ~,
\label{eqP1}\\&&\non\\
0&=&
-{i\over 4}D^2\pa^{\alpha\dot{\alpha}}H_{\alpha\dot{\alpha}}
-{3\over 16}D^2\Db^2F+a\,m^2F \label{eqPb1}~.
\eea
Since
$a\ne 0$ and $m\ne 0$,  
the equations (\ref{eqP1}) and (\ref{eqPb1}) imply 
$\Db_\ad\Bar{F}=D_\a F=0$. Now,  we can use the identity  
${1\over 16}D^2\Db^2+{1\over 16}\Db^2D^2+{1\over 8}D_\a\Db^2D^\a=\Box$, 
in order  to rewrite eqs. (\ref{eqP1}) and (\ref{eqPb1}) as
\bea
0&=&{i\over 4}\Db^2\pa^{\alpha\dot{\alpha}}H_{\alpha\dot{\alpha}}
-3\Box\Bar{F}+a\,m^2\Bar{F}
\,\,\,\, ,\label{eqP2}\\&&\non\\
0&=&-{i\over 4}D^2\pa^{\alpha\dot{\alpha}}H_{\alpha\dot{\alpha}}
-3\Box F+a\,m^2F
\label{eqPb2}\,\,\,\,.
\eea
Next, by applying  ${i\over 4}\Bar{D}^2\pa^{\alpha\dot{\alpha}}$ to eq. 
(\ref{eqH1}) and making  use of eq. (\ref{eqP2}), we arrive at 
\beq
\Box\Big( {2a\over 3}-3\Big)\Bar{F}+a\,
m^2 \Bar{F}=0\,\,\,\,.
\eeq
Choosing  $a\equiv{9\over 2}$ gives $\Bar{F}=0=F$ on the mass shell. 
After that, eqs. 
(\ref{eqP2}) and (\ref{eqPb2}) 
give
$\Pi^L_0 H_{\alpha\dot{\alpha}}=0$. 
${}$Finally, by applying  to
equation (\ref{eqH1}) 
respectively the projectors 
$\Pi^L_{{1\over 2}}$, $\Pi^T_{{1\over 2}}$ and $\Pi^T_1$ we find
\beq
\Pi^L_{{1\over 2}}H_{\a\ad}=\Pi^T_{{1\over 2}}H_{\a\ad}=\Pi^T_1H_{\a\ad}=0
~.
\eeq
The only non-zero projected component is 
$\Pi^T_{3\over 2}H_{\a\ad}$ which is now equal to $H_{\a\ad}$ and, once 
simplified equation (\ref{eqH1}), results to satisfy the Klein-Gordon 
equation. 
All the previous relations imply that on-shell $H_{\a\ad}$, satisfy
equations (\ref{32irrepsp}),  and it describes the irreducible massive 
superspin-$3/2$ multiplet.
The final action is:
\bea
\tilde{S}^{({\rm I})}_{\rm mass} [H, F] 
&=&
S^{({\rm I})}[H, \s]
-{1\over2} m^2
\int d^8z \,
\Big\{ H^{\a\ad} H_{\a\ad}
-9 
F\Bar{F}
 \Big\}~,
\label{IAcomplex2}
\eea
with $\s$ expressed via $F$ according to (\ref{constrIAcomplex}).

The model constructed, eq. (\ref{IAcomplex2}),  
can be related to  that given in  (\ref{IA}). Indeed, let us  
consider the following nonlocal field redefinition (compare with 
\cite{BuchbinderKuzenko}):
\bea
F = - {1 \over 4} \, {D^2 \over\Box}\, \s +\varphi +
{ 1 \over \sqrt{2} } (\cu + {\rm i} \,\cv )~. 
\label{fr}
\eea
Here $\s$ and $\varphi$ are chiral scalars, 
${\Bar D}_\ad \s = {\Bar D}_\ad \varphi =0$, 
while  $\cu$ and $\cv$ are real linear superfields, 
\bea
{\Bar D}^2 \cu = {\Bar D}^2 \cv = 0~, \qquad \quad  
{\Bar \cu} = \cu~, \quad {\Bar \cv} =\cv~.
\eea
We then have
\bea
\int d^8z \,F\Bar{F}
= {1\over 2} \int d^8z  \, (P^2 +V^2)~,
\eea
where 
\bea
P= - {1 \over 4} \, {D^2 \over\Box}\, \s 
- {1 \over 4} \, {{\Bar D}^2 \over\Box}\, {\Bar \s}
+ \cu ~, \qquad 
V = \varphi + {\Bar \varphi} + \cv
\eea
are unconstrained real superfields.
It is obvious that 
$\s=-{1\over 4}\Db^2 F = -{1\over 4}\Db^2 P$.
Let us implement the field redefinition 
(\ref{fr}) in the action  (\ref{IAcomplex2}). 
This gives 
\be
\tilde{S}^{({\rm I})}_{\rm mass} 
[H,F] 
= 
S^{({\rm I})}_{\rm mass} 
[H,P]
+{9 \over 4}m^2 \int d^8z \,V^2~.
\label{IAcomplex3}
\ee
Since $V$ is unconstrained and appears in the action without derivatives, 
it can be integrated out.  This amounts to
 setting to zero the second term in (\ref{IAcomplex3}).

It is worth saying a few more words about the two solutions, 
eqs.  (\ref{constrIAreal}) and   (\ref{constrIAcomplex}),
to the chirality constraint  in terms of unconstrained superfields.
Parametrization (\ref{constrIAcomplex}) 
for the chiral compensator is known to lead to the standard auxiliary fields
of minimal supergravity $(S,P, A_{a})$. If one instead parametrizes 
$\s$ according to (\ref{constrIAreal}), the set of auxiliary fields 
becomes $(S, C_{abc}, A_a)$. This set includes a gauge three-form
$C_{abc}$, instead of the scalar $P$. 
The latter actually occurs as a gauge-invariant field strength 
associated with $C_{abc}$.
It perhaps is worth noting that this three-form
in four dimensions though non-dynamical may be
regarded as the truncation of the well-known
similar dynamical field that occurs in 11D
supergravity and M-theory.

The theory with action (\ref{IAcomplex2}) can be used to 
construct a dual formulation, in a manner similar to 
the approach advocated in \cite{BGKP}.
Instead of imposing eq. (\ref{constrIAcomplex})  
as a kinematic constraint, one can generate it as an equation 
of motion by means of the introduction of  
an unconstrained complex Lagrange 
multiplier $Y$. Consider the following auxiliary action: 
\bea
S=
S^{({\rm I})} [H, \s]
&+&
\int d^8z \, \Big[
-{1\over 2}m^2H^{\alpha\dot{\alpha}}H_{\alpha\dot{\alpha}}
+{9\over 2}m^2F\Bar{F}
\non\\
&&
\,\,\,\,\,\,\,\,\,\,\,\,\,\,
+3m\,Y\Big({1\over 4}\Db^2F+\s\Big)
+3m\,\Bar{Y}\Big({1\over 4}D^2\Bar{F}+\sba\Big)\Big]~.
\label{dual1}
\eea
Here $\s$ is a chiral superfield unrelated to $F$.
Varying  $Y$ and $\Bar{Y}$ enforces the constraints 
(\ref{constrIAcomplex}), and then 
we are clearly back to 
(\ref{IAcomplex2}). 
On the other hand, if 
we integrate out $F$ and $\Bar{F}$ using their equations of motions 
\bea
{3\over 2}mF+{1\over 4}D^2\Bar{Y}=0 ~, \qquad 
{3\over 2} m\Bar{F}+{1\over 4}\Db^2Y=0~,
\eea
and introduce 
the chiral superfield 
$\c=-{1\over 4}\Db^2Y$
and its conjugate $\Bar{\chi}=-{1\over 4}D^2\Bar{Y}$, 
we arrive at the following dual action
\bea
S&=&
\int d^8z  \, \Big[
H^{\alpha\dot{\alpha}}\Box
\Big({1\over 2}\Pi^T_{{3\over 2}}-{1\over 3}\Pi^L_0\Big)
H_{\alpha\dot{\alpha}}
-i(\sigma-\Bar{\sigma})\pa^{\alpha\dot{\alpha}}H_{\alpha\dot{\alpha}}
-3\sigma\Bar{\sigma}\non\\
&&\,\,\,\,\,\,\,\,\,\,\,\,\,\,\,\,\,\,
-{1\over 2}m^2H^{\alpha\dot{\alpha}}H_{\alpha\dot{\alpha}}
-2\,\Bar{\chi}\chi\Big]
+3m \int d^6z\,   \chi\,\s
+3m \int d^6 {\bar z} \,\Bar{\chi}\,\sba ~.
\label{ICC}
\eea
This dynamical system  is quite interesting in its own rights.
Unlike the massive models (\ref{IA})
and (\ref{IAcomplex2}),  the above formulation involves
only the chiral compensator of old minimal supergravity, 
and not its gauge potential.  
The mass generation becomes possible due to the presence of 
a second chiral superfield.
In a  sense,
one can also interpret (\ref{ICC}) as a coupling of the gravitational superfield
to a massive $\cn=2$ hypermultiplet.

The explicit structure of action (\ref{ICC})  explains
why all attempts have failed to construct a Lagrangian formulation 
for the massive superspin-3/2 multiplet solely in terms of the old minimal 
supergravity prepotentials $H_a$, $\s$ and $\Bar \s$.

\subsection{Massive extension of non-minimal supergravity}

Up to now we have considered massive extensions of old minimal 
and new minimal supergravity. Here we would like to address the problem 
of deriving a massive extension of linearized non-minimal supergravity 
\cite{CompensatoriConformi,GatesJV,SUPERSPACE,BuchbinderKuzenko}. 
This goal can be achieved by performing 
a different duality transformation starting from (\ref{IAcomplex2}).

In the action (\ref{dual1}), the superfield $\s$ is chiral by construction.
Enforcing the equation of motion for $Y$ constrains $F$ to be related 
to $\s$ according to (\ref{constrIAcomplex}). 
Clearly, in analogy with the action (\ref{dualIAIB}), we can
actually remove the chirality constraint imposed on $\s$ 
and choose this superfield to be unconstrained complex 
off the mass shell.
Note also that, in such a setting,
we can write a more general action 
that should reduce to (\ref{dual1}) once $\s$ 
is constrained to be chiral, namely\footnote{One can actually consider
even more general action by letting the parameter $a$ and $b$ \\$~~~~~$ to be complex,
$a(S\,D^\a{\Bar D}^\ad H_{\a \ad }-\Bar S\,{\Bar D}^\ad  D^\a H_{\a\ad})~\to~
(aS\,D^\a{\Bar D}^\ad H_{\a \ad }-\Bar{a}\Bar S\,{\Bar D}^\ad  D^\a H_{\a\ad})$\\$~~~~~$
and $b(S^2 + \Bar{S}^2) ~\to ~(b S^2 +{\Bar b}\, {\bar S}^2)$. For simplicity, we 
restrict our consideration to
the \\$~~~~~$ case $a=\Bar{a}$ and $b =\Bar b$.}
\bea
S&=&\int d^8z \, \Bigg[~
H^{\alpha\dot{\alpha}}\Box
\Big({1\over 2}\Pi^T_{{3\over 2}}-{1\over 3}\Pi^L_0\Big)
H_{\alpha\dot{\alpha}}\non\\
&&~~~~~~~~~~
+(2a-1)(S-\Bar{S})i\pa^{\alpha\dot{\alpha}}H_{\alpha\dot{\alpha}}
+aS\,D^\a {\Bar D}^\ad H_{\a \ad }  - a\Bar S\,  {\Bar D}^\ad  D^\a H_{\a\ad}
\non\\
&&~~~~~~~~~~
-3 S \Bar{S}-b\,{3\over 2}\,(S^2 + \Bar{S}^2)\non\\
&&~~~~~~~~~~
-{1\over 2}m^2H^{\alpha\dot{\alpha}}H_{\alpha\dot{\alpha}}
+{9\over 2}m^2F\Bar{F}
\non\\
&&~~~~~~~~~~
+3m\,Y\Big({1\over 4}\Db^2F + S\Big)
+3m\,\Bar{Y}\Big({1\over 4}D^2\Bar{F}+{\Bar S} \Big)\Bigg]\,\,.
\label{dual1a}
\eea
Clearly, varying $Y$ and $\Bar{Y}$ gives 
$S= \s =- \frac{1}{4} \Db^2F $ and the conjugate relation, 
and then 
we are still back to (\ref{IAcomplex}). Instead
if we integrate out $S$, 
$F$ and their conjugates, 
using their equations of 
motion
\bea
F=-{1\over 6m}D^2\Bar{Y}~,&&\quad
0=-S - b \Bar{S}-{(2a-1)\over 3}i\pa_{\a\ad}H^{\a\ad} 
-{a\over 3}\Db_\ad D_\a H^{\a\ad}+ m\Bar{Y},\,\,\,\,\,\,\,\,\,\,\,\,
\eea
which imply
\bea
S &=&
{a\over 6(b+1)}
[D_\a , \Db_\ad ] H^{\a\ad}
+{(a-1)\over 3(b-1)}i\pa_{\a\ad}H^{\a\ad}
+{m\over b^2-1} (b \,Y-
\Bar{Y})~,~~
\eea
we arrive at the following action (defining $\S=mY$
and 
$\chi=-{1\over 4}\Db^2Y$):
\bea
S &=&\int d^8z\Bigg\{
\,\,{1\over 2}\,H^{\alpha\dot{\alpha}}\Box
\Bigg[
\Pi^T_{{3\over 2}}
+ \Big(
{4a^2\over 3(b+1)}-{4(a-1)^2\over 3(b-1)}-{2\over 3}\Big)\Pi^L_0
\non\\
&&\,\,\,\,\,\,\,\,\,\,\,\,\,\,\,\,\,\,\,\,\,
-{4(a-1)^2\over 3(b-1)}\,\Pi^L_{1\over 2}
-{4a^2\over b+1}\,\Pi^T_{{1\over 2}}
\Bigg]H_{\alpha\dot{\alpha}}
\non\\
&&\,\,\,\,\,\,\,\,\,\,\,\,\,\,\,\,\,\,\,\,\,
-{1\over 2}m^2H^{\alpha\dot{\alpha}}H_{\alpha\dot{\alpha}}
+\Big({2a-b-1\over b^2-1}\Big)(\S-\Sb)i\pa^{\a\ad}H_{\a\ad}
\non\\
&&\,\,\,\,\,\,\,\,\,\,\,\,\,\,\,\,\,\,\,\,\,
+{a\over b+1}\,H^{\a\ad}(D_\a\Db_\ad\S-\Db_\ad D_\a\Sb)
+{3b\over 2(b^2-1)}(\S^2+\Sb^2)
\non\\
&&\,\,\,\,\,\,\,\,\,\,\,\,\,\,\,\,\,\,\,\,\,
-{3\over b^2-1}\,\S\Sb
-2\chi\Bar{\chi}\,
\Bigg\}\,\,,
\label{CNMsugra}
\eea
with the dynamical variables $\chi$ and $\S$
constrained as follows:
\bea
\Db_\ad\chi\,=\,0~, \qquad 
-{1\over 4}\Db^2\S\,=\,m\,\chi ~.
\label{constrCNMsugra}
\eea
These constraints describe a chiral--non-minimal (CNM) doublet \cite{CNM}
(see also \cite{CNM0404222} for recent results on the quantum beahviour of CNM 
multiplets).   
We  have thus constructed  
a CNM formulation for massive supergravity. 
In particular, it is easy  to see that the choice 
\bea
a=-{1\over 2}~, \qquad
b=-{3 n-1\over 3n+1}
\eea
corresponds to
\bea
S^{\rm NM}_{\rm mass}&=&\int d^8z\Bigg[
-{1\over 16}\,H^{\alpha\dot{\alpha}}D^\b\Db^2D_\b H_{\alpha\dot{\alpha}}
+{n+1\over 8n}\,(\pa_{\a\ad}H^{\alpha\dot{\alpha}})^2\non\\
&&\,\,\,
+{n+1\over 32}\,([D_\a,\Db_\ad]H^{\alpha\dot{\alpha}})^2
-{(n+1)(3n+1)\over 4n}\,iH^{\alpha\dot{\alpha}}\pa_{\alpha\dot{\alpha}}
(\S-\Bar{\S})\non\\
&&\,\,\,
-{3n+1\over 4}\,H^{\alpha\dot{\alpha}}(D_\a\Db_\ad \S-\Db_\ad D_\a\Bar{\S})
+{(3n+1)^2\over 4n}\,\Bar{\S}\S\non\\
&&\,\,\,
+{9n^2-1\over 8n}\,(\S^2+\Bar{\S}^2)
-{1\over 2}m^2H^{\a\ad}H_{\a\ad}
-2\,\Bar{\chi}\chi\Bigg]
\,\,\,,
\label{mNM}
\eea
with $\chi$ and $\S$ constrained as in  (\ref{constrCNMsugra}).
This model can be recognized  to be the desired
massive extension of non-minimal supergravity (\ref{NM}). 
Since we have derived (\ref{mNM}) 
by applying a  superfield duality transformation to (\ref{IAcomplex2}), 
the two theories are equivalent and describe the massive 
superspin-$3/2$ multiplet.

One can also obtain 
the non-minimal formulation (\ref{mNM}), (\ref{constrCNMsugra}) 
using a slightly different path.
The linearized supergravity actions (\ref{oldM}) and  (\ref{NM})
are known to be dual to each other.
The duality proceeds, say,  by making use of  the auxiliary  action
\bea
S[H,\S,\s]&=&
S^{\rm NM}[H,\S]
-3\int d^8z\Big[\,\s\S+\sba\Sb\,\Big]~,
\label{NMdualI}
\eea
where  $\s$ is chiral and $\S$ is  unconstrained. 
Varying $\s$ in (\ref{NMdualI}) makes $\S$ linear
and then we are back to 
linearized non-minimal supergravity described by (\ref{NM}). 
Instead,  integrating out $\S$ and $\Bar{\S}$ leads to 
the old minimal supergravity  action (\ref{IA}).

Now, in order to find a massive extension of  (\ref{NM}), 
we can start directly from the above action (\ref{NMdualI}) 
extended in the  
following way
\bea
S_m[H,\S,\s,\chi]&=&S^{\rm NM}[H,\S]\non\\
&&
+\int d^8z\Big[-3(\s\S+\sba\Sb)
-{1\over 2}m^2H^{\a\ad}H_{\a\ad}
-2\,\Bar{\chi}\chi\Big]
\non\\
&&
+\,3m\int d^6z\ \chi\,\s
+\,3m\int d^6\Bar{z}\ \Bar{\chi}\,\sba
\,\,\,,
\eea
where $\chi$ is chiral. 
Integrating out $\S$ and $\Bar{\S}$, 
we arrive at the action (\ref{ICC}) which is  
known  to be dual to linearized old minimal supergravity (\ref{IA}). 
Instead, integrating out $\s$ and $\sba$ 
leads to the massive non-minimal 
formulation (\ref{mNM}), (\ref{constrCNMsugra}).

Let us analyse the compensator sector of  (\ref{mNM})
which is obtained by setting $H_a =0$.  Up to a sign, 
it corresponds to  a massive chiral--non-minimal (CNM)
multiplet \cite{CNM}. 
Such a multiplet can be viewed 
as the mechanism to generate a mass 
for the complex linear superfield $\S$ in the presence of a chiral superfield
$\chi$ by means of a consistent deformation of the off-shell constraint: 
${\Bar D}^2 \S = 0 \to  {\Bar D}^2 \S =-4 m \chi$.
The CNM multiplet is know to be dual to a pair of chiral superfields
having a Dirac mass term
of the form $(m\int d^6z\,\s\chi+ {\rm c.c.})$; this multiplet is sometimes called
chiral-chiral (CC). The compensator sector of (\ref{ICC}) is clearly described
by a CC multiplet. It is worth pointing out that  CNM multiplets are ubiquitous
in $\cn=2$ supersymmetry in the framework of projective superspace;
see \cite{ProjectiveSuperspace1,ProjectiveSuperspace2,ProjectiveSuperspaceGR,ProjectiveSuperspaceGK} for references on 4D projective superspace and 
also \cite{ProjectiveSuperspaceED} for extensions to 5 and 6 dimensions.

We close the section by observing that 
in the CC and CNM  massive supergravity formulations developed, 
%respectively 
see eqs. (3.19) and (3.23)--(3.26) respectively, 
the mass parameter $m$ can be 
%freely considered complex. 
easily promoted to become  complex.
This is different from the 
previously known formulation described in the Appendix, 
 and could be a relevant
property 
%in 
when trying to extend these multiplets to extra-dimensions 
in particular for the $D> 5$ case.

\setcounter{equation}{0}
\section{Discussion}

In this work, we have continued (and hopefully completed) a program 
of the exploration of the structure of massive linearized
4D  $\cal N$ = 1 superfield supergravity models.
One of the points of this continued effort is to establish
a number of benchmarks for other purposes.

First, it is known that closed superstring theories
and M-theory, when truncated to four dimensions,
must possess massive spin-2 (and higher) multiplets
in a low-energy effective action.  Thus our effort
is part of the long-term program begun in references
\cite{BGLP1,GKP} to gain a systematic understanding first
of the massive superspin-3/2 
system and later all
of arbitrary 4D $\cal N $ = 1 higher spin multiplets.

Second, massive theories are also interesting to
study as a step toward the realization of higher
values of D as shown in the work of \cite{Linch2003,Gates2003}.  
There a successful approach was given in the case of 
5D supergravity.  However, to date no successful 
extension of this construction is known for higher
values of D.  Thus, this present effort also is a
probe for furthering this program of constructing
(at least) linearized versions of all higher D
supergravity theories in terms of 4D, $\cal N$ = 1
superfields.

We have presented the first successful description
of the massive version of linearized non-minimal 4D
$\cal N$ = 1 superfield supergravity.  As well we
have obtained results that show signs of $\cal N$
= 2 supermultiplet being very relevant to this
course of study.  This result is important in
a way that may also open a new view of the five-dimensional
theory.  The version of the 5D theory constructed
in \cite{Linch2003} only possesses 5D Lorentz invariance
{\it {on-shell}}.  This is manifest in the fact
that though the physical spinors in the work by
Linch, Luty and Phillips \cite{Linch2003}
are proper 5D spinors,
the auxiliary spinors in the work are not.  The
supergravity multiplet in this work is described
by the old minimal supergravity theory given in
\cite{VariantSG}.  This possesses no {\it {auxiliary}} spinors.
A distinguishing point of our present work is
that by describing the supergravity multiplet
in terms of non-minimal supergravity, there 
opens the possibility to contruct a 5D extension
where the auxiliary spinors also describe {\it
{off-shell}} 5D spinors.

Conceptually, the structure of the massive non-minimal action (\ref{mNM})
differs considerably  from the massive minimal models 
(\ref{IA}), (\ref{IIA}) and (\ref{IIIA}), in the sense that (\ref{mNM})
does not involve  any mass term
for the gauge spinor  potential $\Psi_\a$ 
associated with  the non-minimal compensator
$\Bar \S=D^\a \Psi_\a$. To explain this feature, 
let us consider the massive extensions of old minimal supergravity
(\ref{IA}), (\ref{IAcomplex2}) and (\ref{ICC}). These three actions
look identical in the sector involving the gravitational superfield.
Their parts involving the compensators only, obtained by setting
$H_a=0$, look quite different. Nevertheless,  they all share one 
important common feature:
on the mass shell, they describe two free massive superspin-0 multiplets.
The same property holds for the compensator sector of 
the non-minimal action (\ref{mNM}). That is, it describes a free massive 
$\cn=2$ hypermultiplet, or two free massive 
$\cn=1$ superspin-0 multiplets. Let us now introduce 
a massive extension for the compensator part of (\ref{NM}).
This is as follows \cite{BuchbinderKuzenko}: 
\bea
S&=&- \int d^8z\Big[ \S\,{\Bar \S} + {\z \over 2} (\S^2 +{\Bar \S}^2)
+ 2m ( \Psi^\a \Psi_\a + {\Bar \Psi}_\ad {\Bar \Psi}^\ad ) \Big]~,
\label{masscl}
\eea
with $\z$ a parameter. 
Unlike the compensator sector of (\ref{mNM}), 
this action describes a single 
superspin-0 multiplet, since the equations of motion imply 
\be
-{1\over 4} D^2 \S + m {\Bar \S}=0~.
\ee
As a result, the action (\ref{masscl}) can not be used for 
generating a massive extension of linearized non-minimal
supergravity. It is worth pointing out that  
in the massless case, the parameter $\z$ can take arbitrary values 
except $ \pm 1$ \cite{CNM}. In the massive case, no restriction 
on $\z$ occurs, since the corresponding term in (\ref{masscl}) 
can be completely removed by a field redefinition 
$\Psi_\a \to \Psi_\a + (\l /m) D^2 \Psi_\a$, with $\l$ a parameter.

To conclude this paper, we would like to comment upon
a subtle property of the massless action (\ref{NM}) in respect to the 
classification of linearized supergravity models given in \cite{GKP}.
The linearized action for non-minimal supergravity is 
defined for $n \neq -1/3,0$. Looking at the second form for 
the action  (\ref{NM}),  in terms of the superprojectors,
one clearly sees that the case $n=-1$ is very special.
In this and only this case, the action involves only three superprojectors. 
The latter feature appears to be in a seeming contradiction with 
the theorem in \cite{GKP} that there are no irreducible supergravity 
multiplets with three superprojectors in the action. 
Fortunately, this 
contradiction
can be readily resolved if one recalls
the structure of the linearized gauge transformations  
in non-minimal supergravity \cite{BuchbinderKuzenko}: 
\bea
\d H_{\a \ad} &=& {\Bar D}_\ad L_\a - D_\a {\Bar L}_\ad~, \non \\
\d \S&=& -{1\over 4} {n+1\over 3n+1} \,{\Bar D}^2 D^\a L_\a 
- {1\over 4} {\Bar D}_\ad D^2 {\Bar L}^\ad~,
\eea
with $L_\a$ an unconstrained gauge parameter. As may be  seen, 
the gauge freedom allows one to completely gauge away  the complex linear 
compensator $  \S $ provided $n\neq -1$. 
This is no longer true for $n\neq -1$
(in which case 
the compensator can be gauged away on the mass shell only).\footnote{This
property of $n=-1$ supergravity 
is generic within
the so-called gauge transversal formulation for massless multiplets of 
half-integer superspin $Y\geq 3/2$ \cite{KSP}. 
The transversal series terminates at $Y=3/2$ at the $n=-1$
formulation for non-minimal supergravity.}
On the other hand, the classification given in \cite{GKP} applies
to those off--shell realizations for the massless superspin-3/2 multiplet,
which can be formulated solely in terms of the gravitational superfield
upon gauging away the compensator(s).

We hope that the present work has brought the topic of massive off-shell
superspin-3/2 multiplets to the same level of completeness as that 
existing for the massive gravitino multiplets \cite{BGKP,massgravitino}.
\\

\noindent
{\bf Acknowledgements:}\\
The work of S.J.G. is support by the U.S. National
Science Foundation under grant PHY-0353301,
the endowment of the John S. Toll Professorship
and the CSPT. The work  of S.M.K. is supported  
by the Australian Research Council and by a UWA research grant.
G.T.-M. is partially supported by INFN, PRIN prot. 2005-024045-004 and the 
European Commision RTN program MRTN-CT-2004-005104.

\begin{appendix}

\setcounter{equation}{0}
\section{Dual actions}

In this appendix, we collect  the dual formulations 
for the massive minimal models given subsection 2.2 
following \cite{BGKP}.

The theory with action $S^{({\rm I})}_{\rm mass} [H, P]$, 
eq, (\ref{IA}),
possesses a dual formulation.
Let us introduce the ``first-order'' action
\bea
S_{Aux} &=&  \int d^8z \, \Big\{
H^{\a\ad}\Box\Big( \fracov 12 \P^T_{3\over 2}- \fracov{1}{3} \P^L_{0}\Big)H_{\a\ad}
-\fracov{1}{2} m^2 H^{\a\ad}H_{\a\ad}  -U \pa^{\a\ad} H_{\a\ad}
\nonumber \\
&& ~~~~~~~~~~~
-\fracov{3}{2} U^2 + \fracov 94 m^2 P^2
+3m V\Big( U + \fracov{i}{4} \Bar D^2 P - \fracov{i}{4} D^2 P \Big)
\Big\}~,
\label{dualIAIB}
\eea
where
$U$ and $V$ are real unconstrained superfields.
Varying $V$ brings us back to (\ref{IA}).
On the other hand, we can eliminate $U$ and
$P$ using their equations of motion.
With the aid of (\ref{id2}),
this gives
\bea
S^{({\rm IB})} [H, P] &=&  \int d^8z \, \Big\{
H^{\a\ad}\Box\Big(\fracov{1}{2} \P^T_{3\over 2}+\fracov{1}{3} \P^L_{1\over 2}\Big)H_{\a\ad}
-\fracov{1}{2} m^2 H^{\a\ad}H_{\a\ad}
\nonumber \\
&&~~~~~~~~~~- \fracov{1}{16} V \{ \Bar D^2 , D^2 \} V
- m V  \pa^{\a\ad} H_{\a\ad}
+\fracov{3}{2} m^2 V^2 \Big\}~.
\label{IB}
\eea
This is one of the two formulations
for the massive superspin-3/2 multiplet
constructed in \cite{BGLP1}.

The theory (\ref{IIA}) also admits a dual formulation.
Let us consider the following ``first-order'' action
\bea
S_{Aux} &=&\, \int d^8z\,
\Big\{H^{\a\ad}\Box\Big(\fracov{1}{2}\P^T_{3\over 2}-\P^T_{1\over 2}\Big)
H_{\a\ad}
-\fracov{1}{2} m^2 H^{\a\ad}H_{\a\ad}
+\fracov{1}{2}\cu [D_\a,\Bar D_{\dot\a}]H^{\a\ad}
+\fracov{3}{2}\cu^2
\nonumber\\
&&\quad - 6m V \Big( \cu - D^\a \chi_\a
-  \Bar D_{\dot\a}\Bar \chi^{\dot\a} \Big) \Big\}
+3m^2 \Big\{  \int d^6z \,  \chi^\a \chi_\a
+{\rm c.c.} \Big\}~,~~~
\eea
in which $\cu$ and $V$ are real unconstrained
superfields. Varying $V$ gives the original action
(\ref{IIA}). On the other hand, we can eliminate
the independent scalar $\cu$ and chiral spinor
$\chi_\a$ variables using their equations of motion.
With the aid of (\ref{id3}) this gives
\bea
S^{({\rm IIB})} [H, V] &=&\int d^8z\,
\Big\{H^{\a\ad}\Box\Big(\fracov{1}{2}\P^T_{3\over 2}- \fracov{1}{3}\P^L_{0}\Big)H_{\a\ad}
-\fracov{1}{2} m^2 H^{\a\ad}H_{\a\ad}
\nonumber \\
&& \quad +mV  [D_\a,\Bar D_{\dot\a}]H^{\a\ad}
-6m^2 V^2\Big\}
-6  \int d^6z \,  W^\a W_\a~,
\label{IIB}
\eea
where $W_\a=-{1\over 4}\Db^2D_\a V$
is the  vector multiplet field strength.
The theory with   action (\ref{IIB}) 
was constructed in \cite{BGKP}.

${}$Finally, to construct a dual formulation for 
the theory (\ref{IIIA}), let us introduce the ``first-order'' action
\bea
S_{Aux} &=& \int d^8z\,
\Big\{H^{\a\ad}\Box\Big(\fracov 12\P^T_{3\over 2}+\fracov{1}{3}\P^L_{1\over 2}\Big)H_{\a\ad}
-\fracov 12 m^2 H^{\a\ad}H_{\a\ad}
+ \cu \pa_{\a\ad}  H^{\a\ad}
+\fracov 32\cu^2
\nonumber \\
&& +3m V \Big( \cu - D^\a \chi_\a
-  \Bar D_{\dot\a}\Bar \chi^{\dot\a} \Big) \Big\}
-9m^2 \Big\{  \int d^6z \,  \chi^\a \chi_\a
+{\rm c.c.} \Big\}~,
\eea
in which $\cu$ and $V$ are real unconstrained
superfields. Varying $V$ gives the original action
(\ref{IIIA}).
On the other hand, we can eliminate
the independent real scalar $\cu$ and chiral spinor
$\chi_\a$  using their equations of motion.
With the aid of (\ref{id2}) this gives
\bea
S^{({\rm IIIB})} [H, V] &=&\int d^8z\,
\Big\{H^{\a\ad}\Box\Big(\fracov{1}{2}\P^T_{3\over 2}- \fracov{1}{3}\P^L_{0}\Big)H_{\a\ad}
-\fracov{1}{2} m^2 H^{\a\ad}H_{\a\ad}
\nonumber \\
&& \quad -mV  \pa_{\a\ad} H^{\a\ad}
- \fracov 32 m^2 V^2\Big\}
+{1\over 2}  \int d^6z \,  W^\a W_\a~,
\label{IIIB}
\eea
with a vector multiplet field strength
$W_\a$.
This is one of the two formulations
for the massive superspin-3/2 multiplet
constructed in \cite{BGLP1}.  

\end{appendix}

\end{document}